\documentstyle[preprint,aps,prb,epsfig]{revtex}

\tightenlines

\begin{document}
\title{Excess number of percolation clusters on the surface of a sphere}
\author{Christian D. Lorenz \thanks{cdl@umich.edu} and Robert M. Ziff\thanks{rziff@umich.edu}}
\address{Michigan Center of Theoretical Physics and Department of Chemical Engineering University of
Michigan, Ann Arbor, MI 48109-2136}

\date{\today}
\maketitle
\begin{abstract}
Monte Carlo simulations were performed in order to determine the excess number of clusters
$b$ and the average density of clusters $n_c$ for the two-dimensional ``Swiss cheese'' continuum
percolation model on a planar $L \times L$ system and on the surface of a sphere.  The excess number
of clusters for the
$L
\times L$ system was confirmed to be a universal quantity with a value $b = 0.8841$ as previously
predicted and verified only for lattice percolation.  The excess number of clusters
on the surface of a sphere was found to have the value $b =
1.215(1)$ for discs with the same coverage as the flat critical system.  Finally, the
average critical density of clusters was calculated for continuum systems $n_c = 0.0408(1)$.

\end{abstract}

\section{Introduction}
\label{sec1}
One of the interesting characteristics of the percolation model is the existence of universal quantities, which are
independent of the microscopic qualities of the system.  Critical exponents and amplitude ratios are
examples of one class of universal quantities that are only dependent on the dimensionality of the
system.  Shape-dependent universal quantites, which depend on the shape of the boundary as well as the
dimensionality of the system, have been identified in the Ising model by Privman and Fisher
\cite{PF}, M\"{u}ller \cite{muller} and Kamieniarz and Bl\"{o}te
\cite{KB}.  In percolation, recent research has been focused on shape-dependent universal
quantities such as the crossing probabilities
\cite{Lang1,Lang2,Aizen,Cardy,Ziff,HA,Huetal1,Huetal2,Watts,Pinson,ZLK,LZ1} and the excess number of
clusters \cite{ZLK,LZ1,ZFA,KZ} .

The excess number of clusters $b$ on a two dimensional system is defined by \cite{ZLK,ZFA}
\begin{equation}
N \sim n_cA + b 
\label{definition}
\end{equation}
as $A \to \infty$, where $N$ is the total number of the clusters in a system of area $A$, and $n_c$ is
the number of clusters per unit area in an infinite system.  In
(\ref{definition}), it is assumed that the system is at criticality, and that there are no boundaries
to the system.  

The average density of clusters $n_c$ is a non-universal quantity, meaning that it depends on the
microscopic qualities of the system and will have different values for different systems.  In
two-dimensional percolation, exact values of $n_c$ have been determined theoretically for bond
percolation on two lattices \cite{TL,Baxetal}.  Also, simulations have provided precise values of
$n_c$ for 2d \cite{ZFA} and 3d \cite{LZ1} lattice percolation. 

 While $n_c$ is a system-dependent
quantity, $b$ is a universal quantity that is independent of the type of percolation but does
depend on shape of the system boundary and the dimensionality of the system.  Previous work has
demonstrated the shape dependence of $b$ for toroidal systems using 
$L
\times L'$ sytems with periodic boundary conditions, for which
$b(r)$ has been found exactly \cite{KZ}.  For an $L \times L$ system, the theoretical
prediction is $b = 0.883576\ldots$ \cite{KZ}, which agrees with the numerical value $b = 0.8835$
\cite{ZFA} found for site and bond percolation on square and triangular lattices.  For 3d
lattice percolation, theoretical predictions for $\tilde{b}$ (the excess number per unit length) do not
exist; however, the universality of
$\tilde{b}$ has been confirmed using numerical values for different $L \times L \times L'$ cubic
systems
\cite{LZ1}.

In the present paper, we are interested in finding $b$ for percolation on the surface of a sphere,
another two dimensional surface with no boundaries, for which however no theoretical prediction
exists.  In order to do this, we use the 2d continuum ``Swiss cheese'' model of percolation, for which
the critical density has recently been found to high accuracy \cite{QTZpreprint}.  In addition to the
spherical systems, we are interested in using this continuum model to detemine
$b$ for $L \times L$ toroidal systems and comparing it to the theoretical prediction.  In order to
find $b$, we must also find $n_c$, the number of clusters per unit area for the
continuum percolation model. We note that continuum percolation on the surface of a sphere (and
hypersphere) has also been studied in a recent publication \cite{jund}, in the context of diffusion on
fractal clusters.

In the following section, we describe the simulations that were used to model the square and
spherical system.  Then we present and summarize our results for the excess number of clusters and the
average density of clusters in these systems.

\section{Method}
\label{sec2}

The basic ``cluster counting'' algorithm that was used for determining the number of clusters in
the $L \times L$ toroidal and spherical continuum systems is identical to the algorithm we used to
study similar problems in lattice percolation.  However, the implementation of this algorithm was quite
different for the two continuum systems.  

The $L \times L$ system with periodic boundary conditions was first divided into
squares of unit area.  Discs, whose radius
$R$ is equal to 0.5, were distributed into each of the unit squares in the plane using a Poisson
function, where the probability
$P_n$ that there are
$n$ particles in a given volume
$V$ is given by
\begin{equation}
P_n = {1 \over n!} (\rho_c V)^n e^{-\rho_c V}.
\label{poisson}
\end{equation}
Here $\rho_c$ is the critical density of discs ($\rho_c = 1.43628$ for 2d continuum percolation of
discs \cite{QTZpreprint}) and
$V = 1$ is the volume of each of the unit squares.  [Note: The critical density of discs is often
referred to as $n_c$ in the continuum percolation literature, but in order to differentiate from the
average density $n_c$, we use $\rho_c$ here.]\  The algorithms in Ref.\cite{numrec} were used to
generate numbers with this distribution.  For each unit square, a random number $n$ was generated and
if $n > 0$, then
$n$ discs were placed within that square.  The $x$- and $y$- coordinates of each disc were stored in
two one-dimensional arrays, which were indexed by the order that the discs were distributed  (i.\ e.,
the first sphere placed is numbered 0, the second is numbered 1,
$\ldots$).  The index of the first and last disc distributed in the square were also stored in two
one-dimensional pointer arrays.  After discs were placed in each unit square, a search was made for
clusters, starting with the first disc placed in the first unit square.  The search checked only the
neighboring eight unit squares, as opposed to the entire system, for discs.  If the distance between
any two discs was less than or equal to 1, then the two were considered to be in the same cluster.  The
coordinates of each disc in the cluster were stored in two one-dimensional list (``growth'') arrays,
which were indexed by the order that the discs were determined to be part of the cluster.  After the
first disc was checked for overlapping neighbors, then subsequent discs on the list arrays were
checked, in the order that they were placed on the list, for overlapping neighbors.  This process was
continued until a cluster stopped growing; at which time, the same search for each of the unchecked
discs in the current square was performed.  After each of the discs in a square were checked, the
search moved to the next square that had unchecked discs remaining.  This cluster search was continued
until all discs within the system were determined to be part of a cluster.  

When simulating the spherical systems, we encountered two areas of difficulty that were not present
when using the $L \times L$ system.  First, we weren't able to divide the surface into smaller
sub-sections because of the difficulty in producing equal area sub-sections.  Instead, the discs
were distributed on the surface of the entire sphere, using (\ref{poisson}) to determine the number,
and the
$\phi$-coordinate and the cosine of the
$\theta$-coordinate for each disc were stored in two one-dimensional arrays. The entire list was
searched for each neighbor check, resulting in a much slower simulation for larger spherical systems
compared to the toroidal system. We began our cluster search algorithm with the first disc that was
placed on the sphere. The ``great circle'' distance between two discs $d$, whose coordinates are
($\theta_1,\phi_1$) and ($\theta_2,\phi_2$), is given by
\begin{equation}
d = \sqrt{2r^{2}(1-\sin \theta_1 \sin \theta_2 \cos(\phi_1 - \phi_2) - \cos \theta_1 \cos \theta_2)},
\label{arclength}
\end{equation}
which is the arc length between two points on the surface of a sphere of radius $r$.  Two discs were
considered to be overlapping if $d
\le 1$.  The cluster search algorithm was
applied to the spherical system until every disc on the sphere was checked.

The second area of difficulty we encountered was in the selection of
a criterion for the critical threshold that is consistant with the criterion used for a $L \times L$
system.  Different choices (density of discs, surface area coverage, etc.) lead to different numbers of
discs. As a result of this ambiguity, we considered two different methods to determine the average
number of discs that were required.  In general, the total number of the discs $M$ is
defined as 
\begin{equation}
 M = {{\overline M A} \over \overline A} 
\label{coverage}
\end{equation} 
where $\overline A$ is the area covered by
each disc, $\overline M$ is the coverage of the discs, and
$A$ is the total area of the system.  For a finite system,  $M = \rho_c A$, where
$\rho_c$ is the critical density of discs.

First, we used ``type 1'' discs, where we chose the number of discs per unit area on the sphere to have
the same value as the $L \times L$ critical system.  Therefore, the number of discs is
$M = \rho_c(4\pi r^2)$, where $r$ is the radius of the sphere and $\rho_c = 1.43628$ is the critical
density of discs for a flat system.    

\begin{figure}
\centerline{\epsfig{file=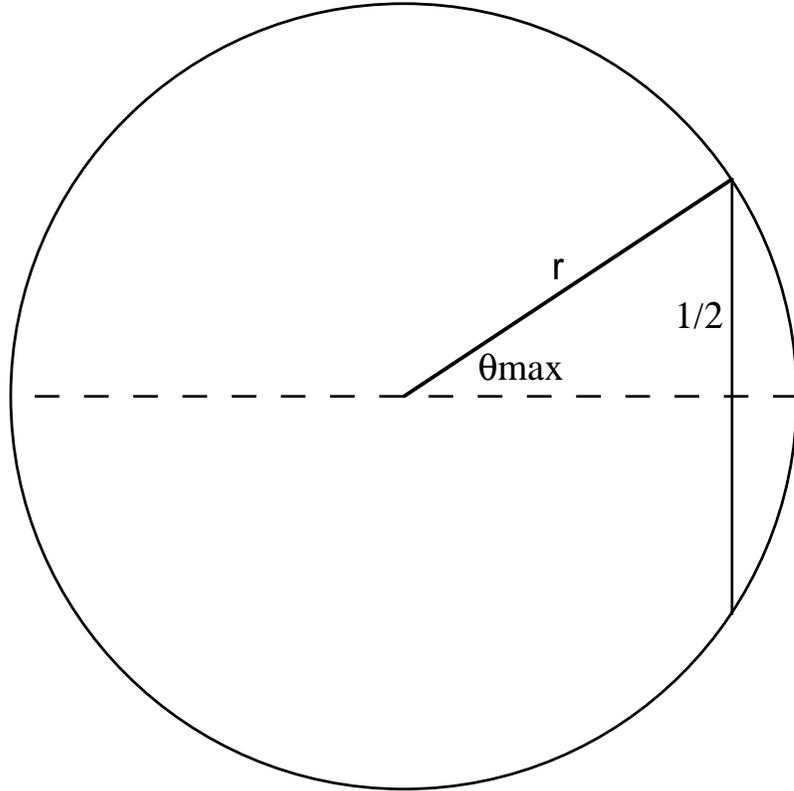, width = 300pt}}
\caption{Drawing of spherical with flat disc on the surface.  The central angle $\theta_{\rm max}$ is
related to the radius $r$ of the sphere by $\sin \theta_{\rm max} = {1/2 \over r}$} 
\label{geo}
\end{figure}

For the second type of discs (``type 2'' discs), we determined the number of discs required to keep
the critical coverage the same as it was for a flat system.  The area covered by each disc is
calculated by
\begin{equation} 
\overline A = \int^{\theta_{\rm max}}_{0} 2\pi r^{2} \sin \theta d \theta = 2 \pi r^{2} (1-\cos
\theta_{\rm max}),
\label{area}
\end{equation}   
 where $r$ is the radius of the sphere, $\theta$ is the angle between the horizontal axis of the
sphere and a point on the surface of the sphere, and $\theta_{\rm max}$ is the angle between the edge
of the disc and the horizontal axis as shown in Fig.\ \ref{geo}.  In this case $\theta_{\rm max} =
1/2r$, therefore, the area covered by each type 2 disc is
\begin{equation}
\overline A = 2 \pi r^{2} (1-\cos 1/2r).
\label{case3area}
\end{equation}
The number $M$ of these discs that would be required to achieve the same coverage as the $L \times L$
system ($\overline M = \rho_c \pi/4$) was determined using (\ref{coverage}), with the result that
\begin{equation}
M = {{\overline M A} \over {\overline A}} = {{\rho_c \pi} \over {2(1-\cos(1/2r))}}
\label{number3}
\end{equation}
where $r$ is the radius of the sphere, where $A = 4 \pi r^2$.

Using these two simulations, we were able to study planar $L \times L$ systems, where $L =$ 8, 16,
32, 64, and 128, and spherical systems with radius $r =$ 5, 6, 10, and 15.  The simulations counted
the number of clusters that existed within the system.  Numerous realizations ($\sim 10^{6}$ for
spherical systems and $\sim 10^{7}$ for planar systems) were averaged over in order to calculate the
density of clusters within each system.     

The random numbers used in these simulations were generated by the four-tap shift-register rule $x_n =
x_{n-471} \oplus x_{n-1586} \oplus x_{n-6988} \oplus x_{n-9689}$, where $\oplus$ is the exclusive-or
operation \cite{Ziff98}.

\section{Results}
\label{sec3}
\begin{figure}
\centerline{\epsfig{file=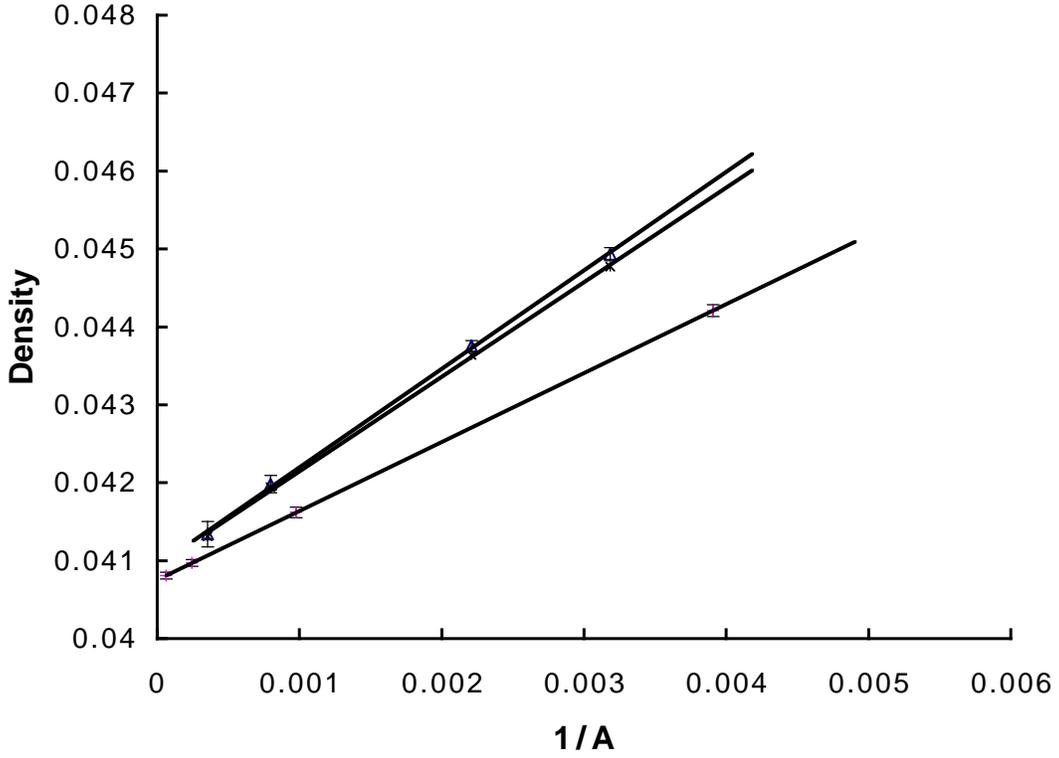, width = 450pt}}
\caption{Plot of $N/A$ vs. $1/A$ for all three systems.  The three sets of data represent the
spherical system with ``type 1'' discs, the spherical
system with ``type 2'' discs, and the planar
$L
\times L$ system, from top to bottom.  In these plots, the $y$-intercept corresponds to the average
density of clusters $n_c$ and the slope is equal to the excess number of clusters $b$.} 
\label{xs}
\end{figure}

The number of clusters $N$ present in a system which has area $A$ is expected to follow
(\ref{definition}).  Using our simulations, we were able to calculate the total number of clusters
present and then determine the overall density of clusters $n = N/A$ for each system size, which by
rearranging (\ref{definition}) is $n = n_c  + b/A + \ldots $, or
\begin{eqnarray}
\label{excessll}
&n = n_c + b/(L^2) + \ldots \quad[L \times L\ {\rm system}] \\
\label{excesssphere}
&n = n_c  + b/(4\pi r^2) + \ldots \quad[{\rm spherical\ system}].  
\end{eqnarray}
Figure \ref{xs} is a plot of $n$ vs.\ $1/A$ for each of the three systems.  By fitting these plots
with equations (\ref{excessll}) and (\ref{excesssphere}), we were able to determine the values of
$b$ and $n_c$ for these systems from the slope and the intercept, respectively, which are summarized
in Table \ref{results}.  

Several observations can be made from these results. First, the average density of clusters for the 
spherical systems and for the $L \times L$ system is, as expected, the same (within numerical
error).  Also, the excess number of clusters
$b$ for the planar
$L
\times L$ continuum system is consistent with the theoretical value ($b = 0.883576 \ldots$) \cite{KZ}
and the simulation  value ($b = 0.8835$) \cite{ZFA} found for the
$L
\times L$ lattice percolation system, which further confirms the universality of this quantity. 
Finally, the values of the excess number of clusters $b$ for the spherical systems are different than
that found for the planar system, which is expected because the shape and topology of the boundary is
different.  However, there is a  difference in the value of $b$ for the two spherical systems, which
is a result of the different definitions for the critical number of discs placed on the surface
of the sphere.  

When the total number of discs $M$ that are placed on the sphere in each case are compared, the
difference in $b$ becomes more understandable: 
\begin{eqnarray}
&M = \rho_c(4\pi r^2) \quad [{\rm type\ 1\ discs}] \nonumber\\
&M = \rho_c (4\pi r^{2}) + \rho_c (\pi/12) +O(1/r^{2} \ldots) \quad [{\rm type\ 2\ discs}].  
\label{comparenumber}
\end{eqnarray}
wherethe second expression follows from a Taylor series expansion of (\ref{number3}).  The first term
in the expression for type 2 discs is exactly the number of type 1 discs present on the surface of the
sphere.  However, the second term shows that  of the order of one more disc is present on the surface
of the sphere than type 1 disc at the same (critical) density. These relatively small differences in
the number of discs present on the sphere can cause a significant difference in the density of
clusters.  In general, about the critical point, one expects the density of clusters to behave as
\cite{SA}
\begin{equation}
n = n_{c} + a_{0}(\rho - \rho_{c}) + a_{1}(\rho-\rho_{c})^{2} + a_{2}(\rho - \rho_{c})^{2-\alpha} +
\ldots
\label{density}
\end{equation}
where $2-\alpha = 8/3$ in two dimensions, $\rho = M/A$, and $A = 4\pi r^2$.  Then the number of
clusters
$N$ is 
\begin{equation}
N = N_c + a_{0}(M - A\rho_c)+a_{1}{{(M-A\rho_c)^{2}} \over A} + \ldots.
\label{numberofclusters}
\end{equation}
Therefore, if $M$ changes by an amount of order 1, then $N$ will also change by an amount of order 1,
which is the same order as the $b$ term in (\ref{definition}).  Thus, the slight differences in the
two definitions of the total number of discs
$M$ --- even though they are asymptotically identical for large $r$ --- lead to non-zero differences
in the value of
$b$.

\section{Discussion of results}
\label{sec4}
The universality of the excess number of clusters $b$ has been demonstrated using 2-d and 3-d lattice
percolation, but it had never been tested with a continuum model.  Our result $b=0.8841$ for the $L
\times L$ planar system is the same as found for the $L \times L$ lattice percolation system
\cite{KZ}, thus it  further confirms the universality of this quantity. 

The excess number of clusters had never been calculated for spherical systems.  By using the
Swiss cheese continuum model, we were able to study percolation on the
surface of sphere.  We found that the excess number of clusters ($b({\rm type\ 1\ discs}) = 1.263$,
$b({\rm type\ 2\ discs}) = 1.215$) is slightly dependent upon the definition of the critical density of
discs for the spherical system.  We believe that the type 2 definition is the most reasonable as it is
based upon the idea that the surface coverage is the same as the planar system, which is consistent
with the fact that the surface coverage is invariant when the $L \times L$ periodic system is deformed
to a torus.  Therefore, we propose the value $b = 1.215(1)$ for the spherical surface.  While
theoretical results exist for $b$ for the $L \times L'$ periodic system, there is no prediction for the
sphere.

\pagebreak
\begin{table}
\caption{Values of the critical average density of clusters $n_c$ and the excess number of clusters
$b$ for two-dimensional ``Swiss cheese'' model.  Numbers in parenthesis represent the error in the
last digit.}
\begin{tabular}{|lll|} 
System&$n_c$&$b$\\ \hline
Planar ($L \times L$) &$0.04075(5) $&$ 0.8841(2)$\\ 
Spherical (``type 1'' discs) &$0.04093(10)$&$1.263(1)$\\
Spherical (``type 2'' discs)& $0.04092(10)$&$1.215(1)$
\end{tabular}
\label{results}
\end{table}

\end{document}